\documentclass{aastex63}
\usepackage{xcolor}
\usepackage{longtable}
\usepackage{amsmath,amssymb}
\usepackage[figuresright]{rotating}
\usepackage{amssymb}
\usepackage{url}

\makeatletter

\newcommand{\Rmnum}[1]{\expandafter\@slowromancap\romannumeral #1@}
\makeatother
\usepackage{color}
\submitjournal{ApJL}

\shorttitle{CRSF in ULX Swift~J0243.6+6124}
\shortauthors{Kong et al.}

\begin{document}

\title{\emph{Insight}-HXMT discovery of the highest energy CRSF from the first Galactic ultra-luminous X-ray pulsar Swift~J0243.6+6124}

\author[0000-0003-3188-9079]{Ling-Da Kong\textsuperscript{*}}
\email{kongld@ihep.ac.cn}
\affil{Key Laboratory for Particle Astrophysics, Institute of High Energy Physics, Chinese Academy of Sciences, 19B Yuquan Road, Beijing 100049, China}
\affil{University of Chinese Academy of Sciences, Chinese Academy of Sciences, Beijing 100049, China}

\author{Shu Zhang\textsuperscript{*}}
\email{szhang@ihep.ac.cn}
\affil{Key Laboratory for Particle Astrophysics, Institute of High Energy Physics, Chinese Academy of Sciences, 19B Yuquan Road, Beijing 100049, China}

\author{Shuang-Nan Zhang\textsuperscript{*}}
\email{zhangsn@ihep.ac.cn}
\affil{Key Laboratory for Particle Astrophysics, Institute of High Energy Physics, Chinese Academy of Sciences, 19B Yuquan Road, Beijing 100049, China}
\affil{University of Chinese Academy of Sciences, Chinese Academy of Sciences, Beijing 100049, China}

\author{Long Ji}
\affil{School of Physics and Astronomy, Sun Yat-Sen University, Zhuhai, 519082, China}

\author{Victor Doroshenko}
\affil{Institut f{\"u}r Astronomie und Astrophysik, Kepler Center for Astro and Particle Physics, Eberhard Karls, Universit{\"a}t, Sand 1, D-72076 T{\"u}bingen, Germany}
\affil{Space Research Institute of the Russian Academy of Sciences, Profsoyuznaya Str. 84/32, Moscow 117997, Russia}

\author{Andrea Santangelo}
\affil{Institut f{\"u}r Astronomie und Astrophysik, Kepler Center for Astro and Particle Physics, Eberhard Karls, Universit{\"a}t, Sand 1, D-72076 T{\"u}bingen, Germany}

\author{Yu-Peng Chen}
\affil{Key Laboratory for Particle Astrophysics, Institute of High Energy Physics, Chinese Academy of Sciences, 19B Yuquan Road, Beijing 100049, China}

\author{Fang-Jun Lu}
\affil{Key Laboratory for Particle Astrophysics, Institute of High Energy Physics, Chinese Academy of Sciences, 19B Yuquan Road, Beijing 100049, China}
\affil{Key Laboratory of Stellar and Interstellar Physics and Department of Physics, Xiangtan University, Xiangtan 411105, Hunan, China}

\author{Ming-Yu Ge}
\affil{Key Laboratory for Particle Astrophysics, Institute of High Energy Physics, Chinese Academy of Sciences, 19B Yuquan Road, Beijing 100049, China}

\author{Peng-Ju Wang}
\affil{Key Laboratory for Particle Astrophysics, Institute of High Energy Physics, Chinese Academy of Sciences, 19B Yuquan Road, Beijing 100049, China}
\affil{University of Chinese Academy of Sciences, Chinese Academy of Sciences, Beijing 100049, China}

\author{Lian Tao}
\affil{Key Laboratory for Particle Astrophysics, Institute of High Energy Physics, Chinese Academy of Sciences, 19B Yuquan Road, Beijing 100049, China}

\author{Jin-Lu Qu}
\affil{Key Laboratory for Particle Astrophysics, Institute of High Energy Physics, Chinese Academy of Sciences, 19B Yuquan Road, Beijing 100049, China}

\author{Ti-Pei Li}
\affil{Key Laboratory for Particle Astrophysics, Institute of High Energy Physics, Chinese Academy of Sciences, 19B Yuquan Road, Beijing 100049, China}
\affil{Department of Astronomy, Tsinghua University, Beijing 100084, China}
\affil{University of Chinese Academy of Sciences, Chinese Academy of Sciences, Beijing 100049, China}

\author{Cong-Zhan Liu}
\affil{Key Laboratory for Particle Astrophysics, Institute of High Energy Physics, Chinese Academy of Sciences, 19B Yuquan Road, Beijing 100049, China}

\author{Jin-Yuan Liao}
\affil{Key Laboratory for Particle Astrophysics, Institute of High Energy Physics, Chinese Academy of Sciences, 19B Yuquan Road, Beijing 100049, China}

\author[0000-0003-4856-2275]{Zhi Chang}
\affil{Key Laboratory for Particle Astrophysics, Institute of High Energy Physics, Chinese Academy of Sciences, 19B Yuquan Road, Beijing 100049, China}

\author{Jing-Qiang Peng}
\affil{Key Laboratory for Particle Astrophysics, Institute of High Energy Physics, Chinese Academy of Sciences, 19B Yuquan Road, Beijing 100049, China}
\affil{University of Chinese Academy of Sciences, Chinese Academy of Sciences, Beijing 100049, China}

\author{Qing-Cang Shui}
\affil{Key Laboratory for Particle Astrophysics, Institute of High Energy Physics, Chinese Academy of Sciences, 19B Yuquan Road, Beijing 100049, China}
\affil{University of Chinese Academy of Sciences, Chinese Academy of Sciences, Beijing 100049, China}

\begin{abstract}

The detection of cyclotron resonance scattering features (CRSFs) is the only way to directly and reliably measure the magnetic field near the surface of a neutron star (NS). The broad energy coverage and large collection area of \emph{Insight}-HXMT in the hard X-ray band allowed us to detect the CRSF with the highest energy known to date, reaching about 146 keV during the 2017 outburst of the first galactic pulsing ultraluminous X-ray source (pULX) Swift~J0243.6+6124. 
During this outburst, the CRSF was only prominent close to the peak luminosity $\sim 2\times10^{39}$ erg s$^{-1}$, the highest to date in any of the Galactic pulsars. 
The CRSF is most significant in the spin phase region corresponding to the main pulse of the pulse profile, and its centroid energy evolves with phase from 120 to 146 keV. 
We identify this feature as the fundamental CRSF, since no spectral feature exists at $60-70$ keV.
This is the first unambiguous detection of an electron CRSF from an ULX. We also estimate a surface magnetic field $\sim1.6\times10^{13}$ G for Swift~J0243.6+6124. 
Considering that the dipole magnetic field strengths, inferred from several independent estimates of magnetosphere radius, are at least an order of magnitude lower than our measurement, we argue that the detection of the highest energy CRSF reported here unambiguously proves the presence of multipole field components close to the surface of the neutron star. Such a scenario has previously been suggested for several pulsating ULXs, including Swift J0243.6+6124, and our result represents the first direct confirmation of this scenario.

\end{abstract}

\section{Introduction}

Ultraluminous X-ray (ULX) sources are bright X-ray sources with apparent luminosity $>10^{39}$ erg s$^{-1}$, which is above the Eddington limit of  a $10$ $M_{\odot}$ black hole (BH) (\citealp{Kaaret2017ARAA}). 
X-ray pulsations detected in ULX sources such as M82 X-2 (\citealp{Bachetti2014Natur}), NGC 7793 P13 (\citealp{Fuerst2016ApJ,Israel2017MNRAS}), NGC 5907 ULX1 (\citealp{Israel2017Sci}), NGC 300 ULX1 (\citealp{Carpano2018MNRAS}), and RX J0209.6-7427 (\citealp{Chandra2020MNRAS}) indicate that some, or even most ULXs are accreting neutron stars (NSs) with a strong magnetic field.
A potential detection of a proton CRSF at $4.5$ keV in the ULX source M51 ULX-8 implies a magnetar-like field $\sim10^{15}$ G (\citealp{Brightman2018NatAs}). 
Such a strong dipole magnetic field prevents the spherization by radiation in its accretion disk even at luminosities $>10^{41}$ erg s$^{-1}$ (\citealp{Basko1976}).

Alternative estimations of the dipole magnetic fields of NSs can be made from the accretion torque which has a close connection with the inner edge of the accretion disk influenced by the NS's magnetosphere. 
However, different accretion torque models might give large measurement disparities of magnetic fields in the range of $10^{10-15}$ G (\citealp{Chen2021JHEAp}).
The estimated lower values ($<10^{13}$ G) in these pulsating ULXs will make them essentially indistinguishable from Be X-ray pulsars, with the exception of their higher luminosities which has been proposed to be a consequence of the strong beaming effect by the disk outflow, when the disk is spherized by radiation pressure (\citealp{King2001ApJ,King2009MNRAS,King2016MNRAS,King2017MNRAS,Koliopanos2017AA,King2019MNRAS,King2020MNRAS}). 
From the relationship between the magnetic field and maximum luminosity (see Figure 5 in \cite{Mushtukov2015MNRAS.454.2714M}), it can be seen that only when the dipole magnetic field is greater than 10$^{13}$ G, the maximum accretion luminosity of magnetized pulsars can be significantly greater than $10^{39}$ erg s$^{-1}$. 
However, from the first discovery of CRSF in Her X-1 to the present (\citealp{Truemper1978ApJ, Staubert2019}), a direct measurement of such a strong magnetic field has not been possible by observing the expected electron CRSF because of the relatively low cutoff energy of the non-thermal emission from NS ULX systems, and the lack of sensitive hard X-ray coverage above about 80 keV of contemporary X-ray telescopes. 
Fortunately, the launch of \emph{Insight}-HXMT in 2017 (\citealp{2014SPIE.9144E..21Z}; \citealp{2020SCPMA..63x9502Z}), which has a broad energy coverage and a large detection area at energies above 100 keV changed the situation.
The two previous records of the highest energy detection of CRSFs with higher than $5\ \sigma$ significance were also both set by \emph{Insight}-HXMT: the $\sim$90 keV fundamental CRSF from GRO~J1008-57 (\citealp{Ge2020ApJ}) and the $\sim$100 keV first harmonic of a CRSF from 1A~0535+262 (\citealp{Kong2021ApJL}; \citealp{Kong2022arXiv220411222K}) proving that \emph{Insight}-HXMT has a strong advantage in detecting high-energy CRSFs.

The transient X-ray source Swift~J0243.6+6124 was discovered on October 3 2017 by the Swift/BAT telescope at a flux level of $\sim$ 80 mCrab before it entered into a giant outburst from 2017 to 2018 (\citealp{Cenko2017}; \citealp{Kennea2017}).
A 9.8 s period pulsation was detected in X-ray observations by Swift/XRT and an eccentric binary system with a $\sim$ 28-day orbital period was revealed by Fermi-GBM (\citealp{Bahramian2017,Kennea2017,Ge2017,Doroshenko2018}).
Optical observations assigned a Be star as the companion (\citealp{Kouroubatzakis2017,Kennea2017,Stanek2017,Yamanaka2017}).
A distance of 6.8 kpc as reported in \cite{Bailer-Jones2018} based on Gaia DR2 parallax measurements of the companion star implies that the source is the first Galactic pulsing ultraluminous X-ray source (pULX).

The magnetic field of Swift~J0243.6+6124 was estimated in various manners, yet no consensus has been reached on its surface magnetic field strength.
The pure timing analyses based on \emph{Insight}-HXMT's high cadence observations reveal two typical luminosities, around which the source is believed to have experienced transitions of different emission modes at the magnetic pole and different accretion modes on the disk (\citealp{Doroshenko2020MNRAS}), implying a dipole surface magnetic field of $\sim10^{12-13}$ G and non-transition to the propeller state in quiescence. In particular, the non-detection of the transition allowed \cite{Tsygankov2018} to put an upper limit of $\sim6\times10^{12}$\,G assuming propeller transition below $\sim6\times10^{35}$ erg s$^{-1}$, which was further reduced to $\le3\times10^{12}$\,G by \citealp{Doroshenko2020MNRAS} based on NuSTAR detection of the pulsations down to luminosities $\sim6\times10^{33}$\,erg\,s$^{-1}$.
As discussed by \cite{Doroshenko2020MNRAS}, such a low dipole magnetic field allows for a coherent interpretation of the observed spin evolution of the source during the outburst, observed power spectral density, and spectral transition to correspond to the transition of the accretion disk to the radiatively dominated state. 
On the other hand, the estimated critical luminosity corresponding to the onset of an accretion column appeared to be more consistent with a much stronger surface magnetic field $\sim10^{13}$\, G (\citealp{Kong2020ApJ, Liu2022MNRAS}). 
\cite{Doroshenko2020MNRAS}, suggesting that it could be associated with either a smaller than expected magnetospheric radius (for a given field), or the presence of strong multipole components dominating the area close to the surface of the neutron star.
Here we report our very statistically significant detection of a CRSF during the 2017 outburst of Swift~J0243.6+6124 observed with \emph{Insight}-HXMT, when the outburst peaked at a luminosity around $2\times10^{39}$ erg s$^{-1}$ which unambigously supports the latter scenario. We also perform detailed phase-resolved spectral analyses, given that the appearance of CRSFs can be closely phased dependent (\citealp{Harding1991, Nishimura2015}).

Section 2 introduces the observations and methods of data reduction. Section 3 shows the details of the timing and spectral results. In Section 4, we discuss our results. The summary and conclusions are presented in Section 5.

\section{Observations and Data reduction}

The Hard X-ray Modulation Telescope (\citealp{2014SPIE.9144E..21Z}; \citealp{2020SCPMA..63x9502Z}), named \emph{Insight}-HXMT, was launched on June 15, 2017. 
The scientific payload includes three collimated telescopes that allow observations in a broad energy band ($1-250$ keV) and with a large effective area at high energies: the High Energy X-ray Telescope (HE: 5000 $\rm cm^2$/20-250 keV, \citealp{2020SCPMA..63x9503L}), the Medium Energy X-ray Telescope (ME: 952 $\rm cm^2$/5-30 keV, \citealp{2020SCPMA..63x9504C}), and the Low Energy X-ray Telescope (LE: 384 $\rm cm^2$/1-10 keV, \citealp{2020SCPMA..63x9505C}. The main Fields of View (FoVs) are $1.6^{\circ}\times6^{\circ}$, $1^{\circ}\times4^{\circ}$, and $1.1^{\circ}\times5.7^{\circ}$ for LE, ME and HE, respectively.
Ascribed to the characteristics of the three detectors, \emph{Insight}-HXMT possesses a broad-band capability ($1-250$ keV), with a large collection area at high energy and without pile-up for bright sources. 

\emph{Insight}-HXMT was triggered with 122 pointing exposures from Oct 7, 2017 (MJD 58033) to Feb 21, 2018 (MJD 58170) which sampled the entire outburst of Swift J0243.6+6124. 
The \emph{Insight}-HXMT Data Analysis Software (HXMTDAS) v2.04, together with the calibration model v2.05, are used for the whole process of data reduction following the recommended screening conditions by the \emph{Insight}-HXMT team. 
In particular, data with elevation angle (ELV) larger than $10^{\circ}$, geometric cutoff rigidity (COR) larger than 8 GeV, and offset for the point position smaller than $0.04^{\circ}$ are used.
In addition, data taken within 300\,s of the South Atlantic Anomaly (SAA) passage outside of good time intervals identified by onboard software have been rejected. 
In view of the conditions of in-flight calibration (\citealp{Li2020JHEAp}), the energy bands considered for spectral analysis are: $2-10$\,keV for LE, $10-40$\,keV for ME, and $30-250$\,keV for HE. 
The backgrounds are estimated with the tools provided by the \emph{Insight}-HXMT team: LEBKGMAP, MEBKGMAP, and HEBKGMAP, version 2.0.9 based on the current standard \emph{Insight}-HXMT background models (\citealp{Liao2020a,Guo2020JHEAp,Liao2020b}). 
We note that the uncertain calibration of an Ag emission line at $\sim22$ keV in the ME band will leave a narrow emission line in the phase-averaged spectrum (see the left panel in Figure~\ref{spectra}), but it fades out due to lower statistics in the phase-resolved spectrum (see the right panel in Figure~\ref{spectra}). 
Considering the current accuracy of the instrument calibration and appropriate $\chi^2$, we include 1$\%$, 0.5$\%$, and 0.5\% systematic errors for spectral analysis for LE, ME, and HE, respectively.
The uncertainties of the spectral parameters are computed using the Markov Chain Monte Carlo (MCMC) method with a length of 10,000 and are reported at a 90\% confidence level.
The XSPEC v12.12.0 software package (\citealp{1996ASPC..101...17A}) corresponding to heasoft v6.29 is used to perform the spectral fitting and error estimation.
We utilize the ftool \textit{grppha} to improve the counting statistic of the spectrum (with group minimum counts of 200) by combining ten exposures from Nov 03 to Nov 08 in 2017 with the \textit{addspec} and \textit{addrmf} tasks. 
The flux, pulse profile, and continuum spectral shape in these ten exposures keep stable at the outburst peak (\citealp{Kong2020ApJ}); it is therefore reasonable to combine them in order to search for any high energy CRSF with the best statistics.
For the HE detector, the background dominates in the high energy band, and 200 counts per bin will ensure high confidence for the net spectrum. In Figure~\ref{spectra} and Figure~\ref{back_plot}, we use ``setplot rebin 30 5'' for a better visual display, which does not affect the fitting result.

\section{Results}

\subsection{Phase 2-D maps}

\begin{figure}
    \centering\includegraphics[width=0.49\textwidth]{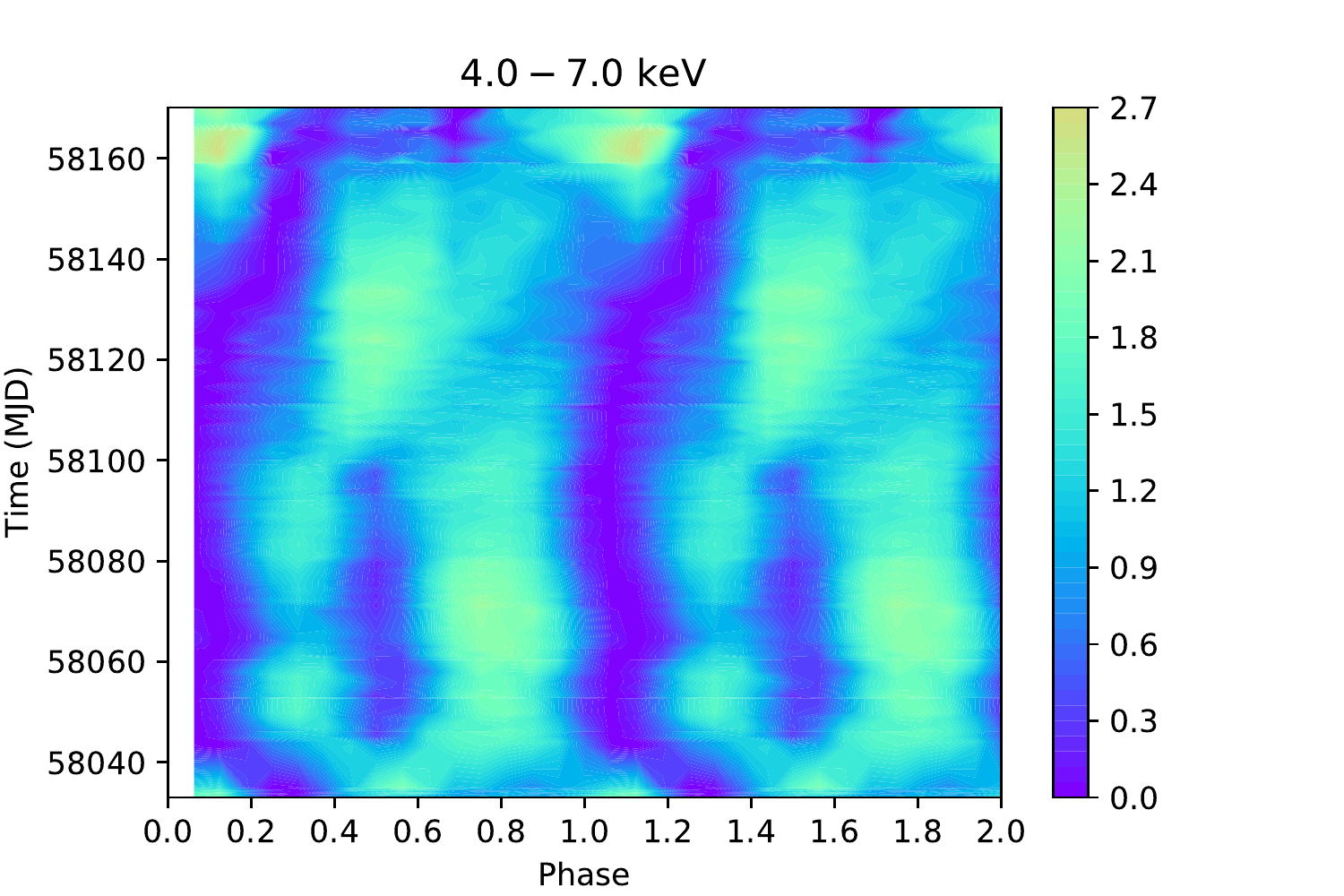}
    \centering\includegraphics[width=0.49\textwidth]{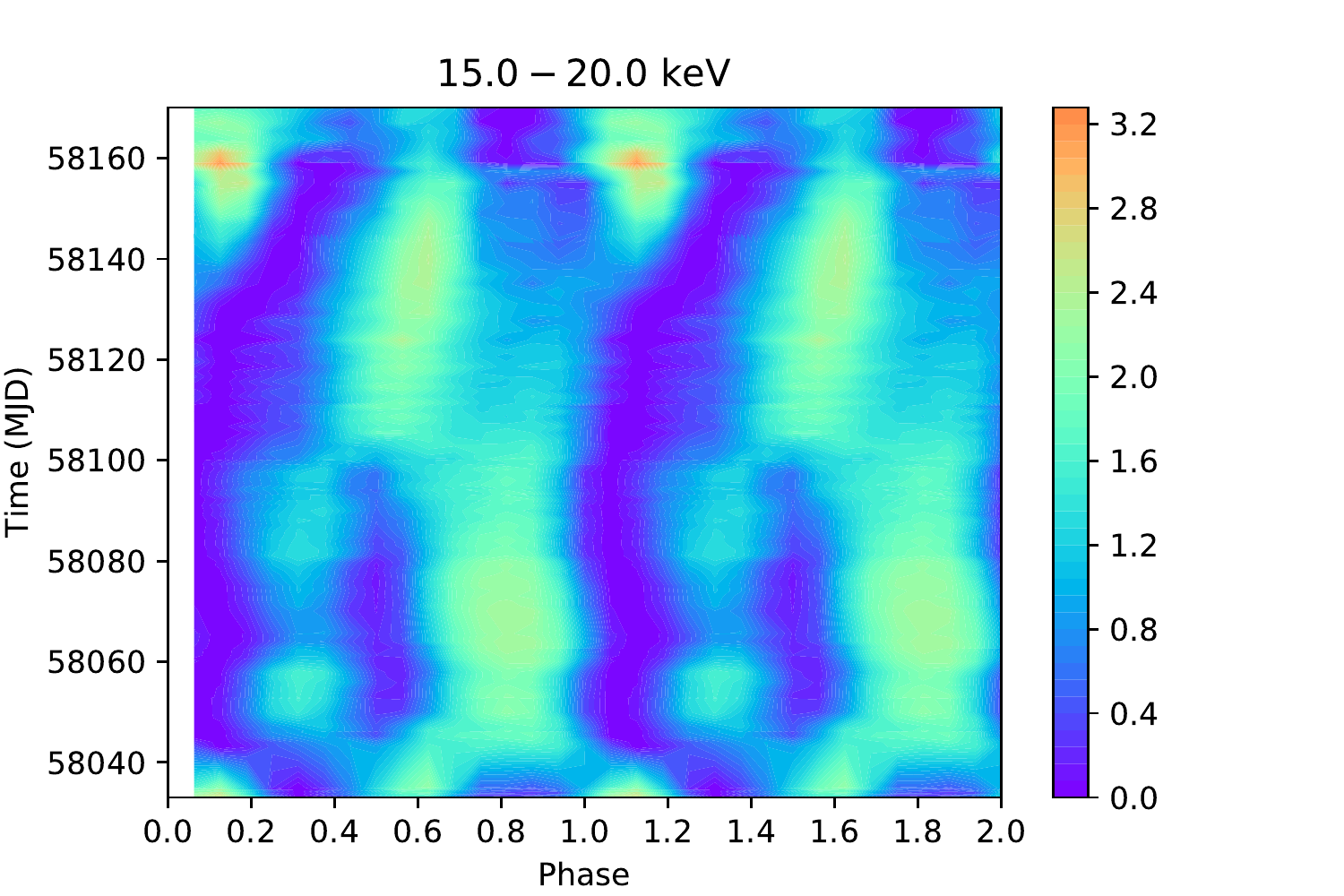}
    \centering\includegraphics[width=0.49\textwidth]{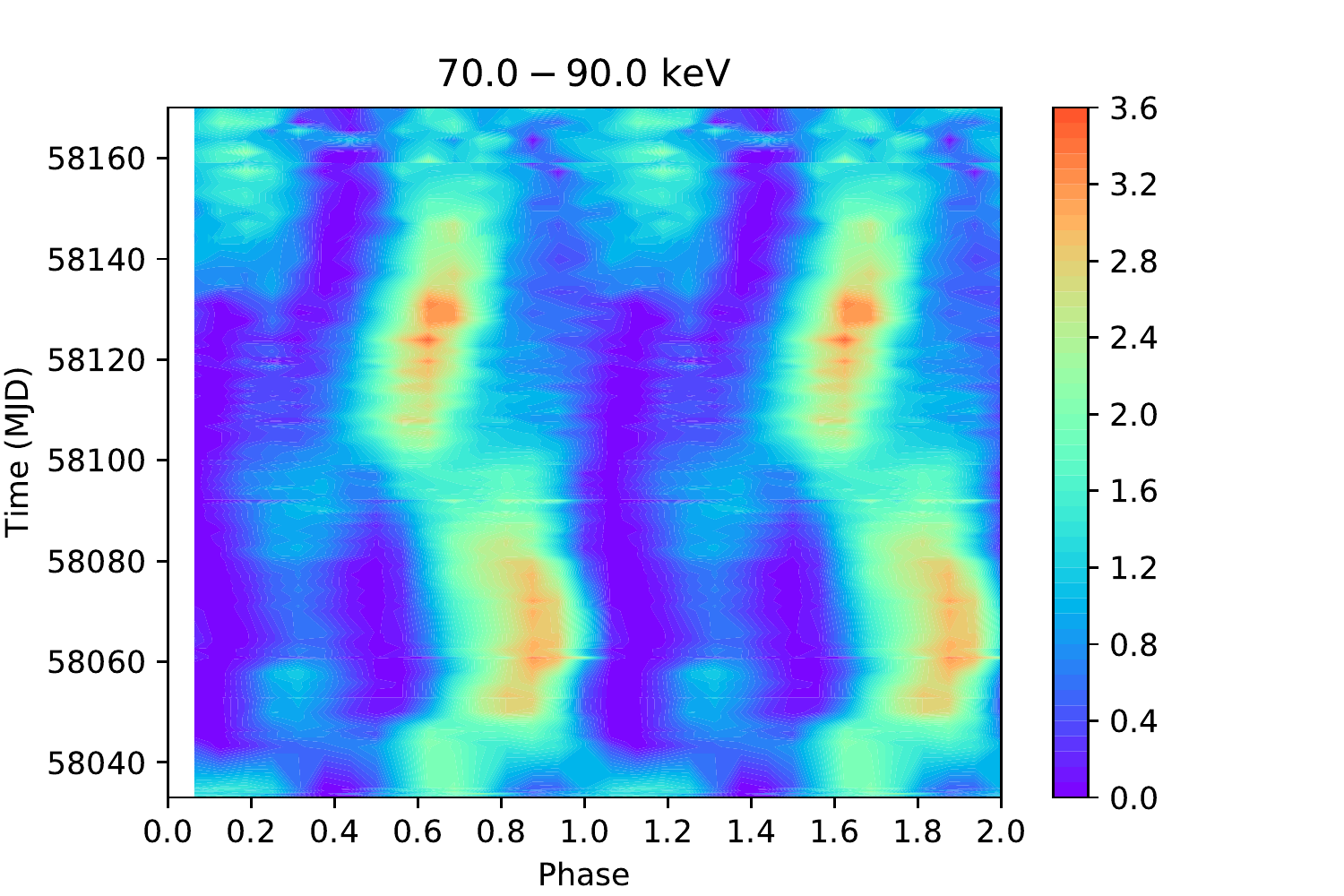}
    \centering\includegraphics[width=0.49\textwidth]{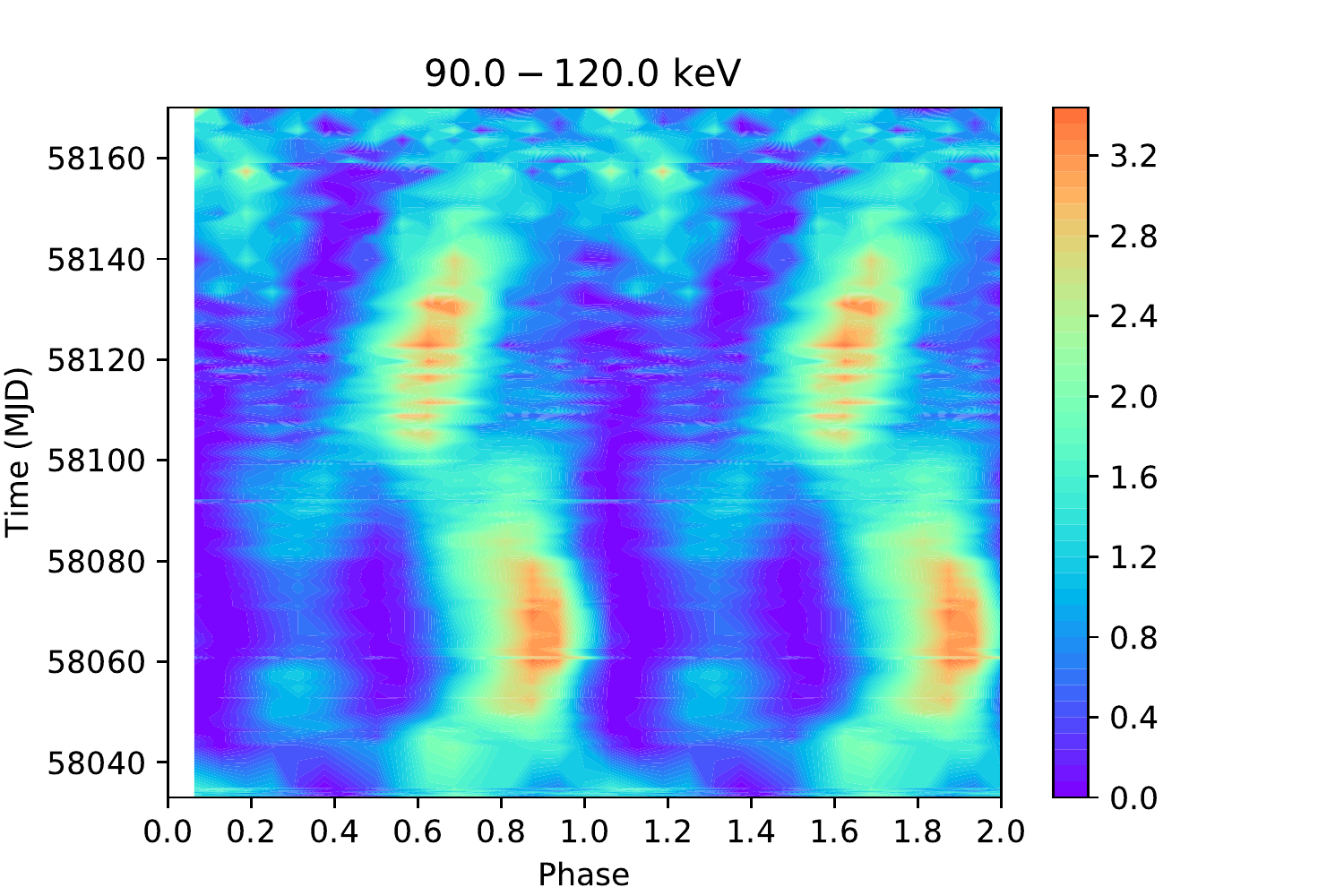}
    \caption{The two-dimension (2-D) maps describe the evolution of the pulse profile with time; the colors plot the values of the normalized pulse values by (Pulse-PulseMin)/(Average count rate-PulseMin), where the PulseMin is the minimum value of the pulse intensity. The pulse profiles are folded from the background-subtracted light curves. The main pulse at phase $0.6-1.0$ is above the mean count rate, and the minor peak at phase $0.2-0.4$ is close to or lower than the average corresponding to the negative value for $70.0-90.0$ keV and $90.0-120.0$ keV.}
    \label{maps}
\end{figure}

The first and foremost step of phase-resolved analysis is to get the correct phase of each photon. After barycentric and binary orbiting corrections, the photon arriving time is converted to the real pulsating time. 
The parameters of the binary orbit ($T_0=58102.97476560854$, $P_{\rm orbit}=27.698899$ day, $e=0.1029$, $ax\sin i=115.531$ light-sec, $\omega=-74.05^{\circ}$) are taken from the website of Fermi/GBM Accreting Pulsar Histories\footnote{https://gammaray.msfc.nasa.gov/gbm/science/pulsars/lightcurves/swiftj0243.html}, where the spin frequencies are available on a daily basis.
With such continuous and intensive monitoring, we can fit the frequency with time by the cubic spline interpolation method. Then the phase-coherent pulse profiles considering the pulse-period changes can be derived by calculating a sequential pulse phase $\phi(t)$, where $\phi_0$ is set to 0 at $t_0= 58,027.499066$ (MJD), which is the epoch of the first Fermi/GBM periodicity detection (see \cite{Sugizaki2020ApJ} for a detailed description of this method). 
In Figure~\ref{maps}, the pulse profiles folded from the background-subtracted light curves are shown in four panels with different energy bands (LE: $4.0-7.0$ keV, ME: $15.0-20.0$ keV, HE: $70.0-90.0$ keV, $90.0-120.0$ keV), and such results are similar to the results in \cite{Doroshenko2020MNRAS}.
From the 2-D maps, we see pulse profile transitions between different shapes at two typical luminosities as reported previously in \cite{Kong2020ApJ} and \cite{Doroshenko2020MNRAS}. 
One is $\sim1.5\times10^{38}$ erg s$^{-1}$ at MJD 58035 and MJD 58130, and another is $\sim4.4\times10^{38}$ erg s$^{-1}$ at MJD 58050 and MJD 58100, probably corresponding to the onset of a pressure dominance of radiation and gas shock, respectively. At the outburst peak, the pulse profile maintains a stable bimodal structure, which allows us to do spectral analysis.
The main pulse in $0.6-1.0$ is above the mean count rate, and the minor pulse in $0.2-0.4$ is close to or lower than the average.
Such phenomena get clearer at higher energies and luminosities: at the peak of the outburst (MJD 58063), the count rate of the main pulse is 4 times that of the mean count rate, while it is far less than 1 for the minor peak (see left panel in Figure~\ref{phase-resolved}). 
Figure~\ref{maps} shows that the best counting statistics above 90 keV comes from the phase region $0.6-1.0$ during the outburst peak (MJD 58060-58065), the data of which can be used to search for high energy CRSFs more reliably.

\subsection{Phase-averaged and phase-resolved spectrum}

We combine ten exposures in 6 days from Nov 3 to 8 for the high-statistic phase-resolved spectral analysis, ensuring that the exposure times of LE, ME, and HE for each phase are more than 2000 s after dividing the spectrum into ten phase intervals.
The pulse profiles in different energy bands are shown in the left panel of Figure~\ref{phase-resolved}. The green dashed lines show the normalized pulse value of 1, and the grey dashed lines ($120-300$ keV) show the normalized pulse value of 0. Significant pulse modulations are observed up to the 120-200 keV energy band with a prominent peak in phase $0.6-1.0$; however, no obvious pulse modulation is visible above 200 keV.

\begin{figure}
    \centering\includegraphics[width=0.49\textwidth]{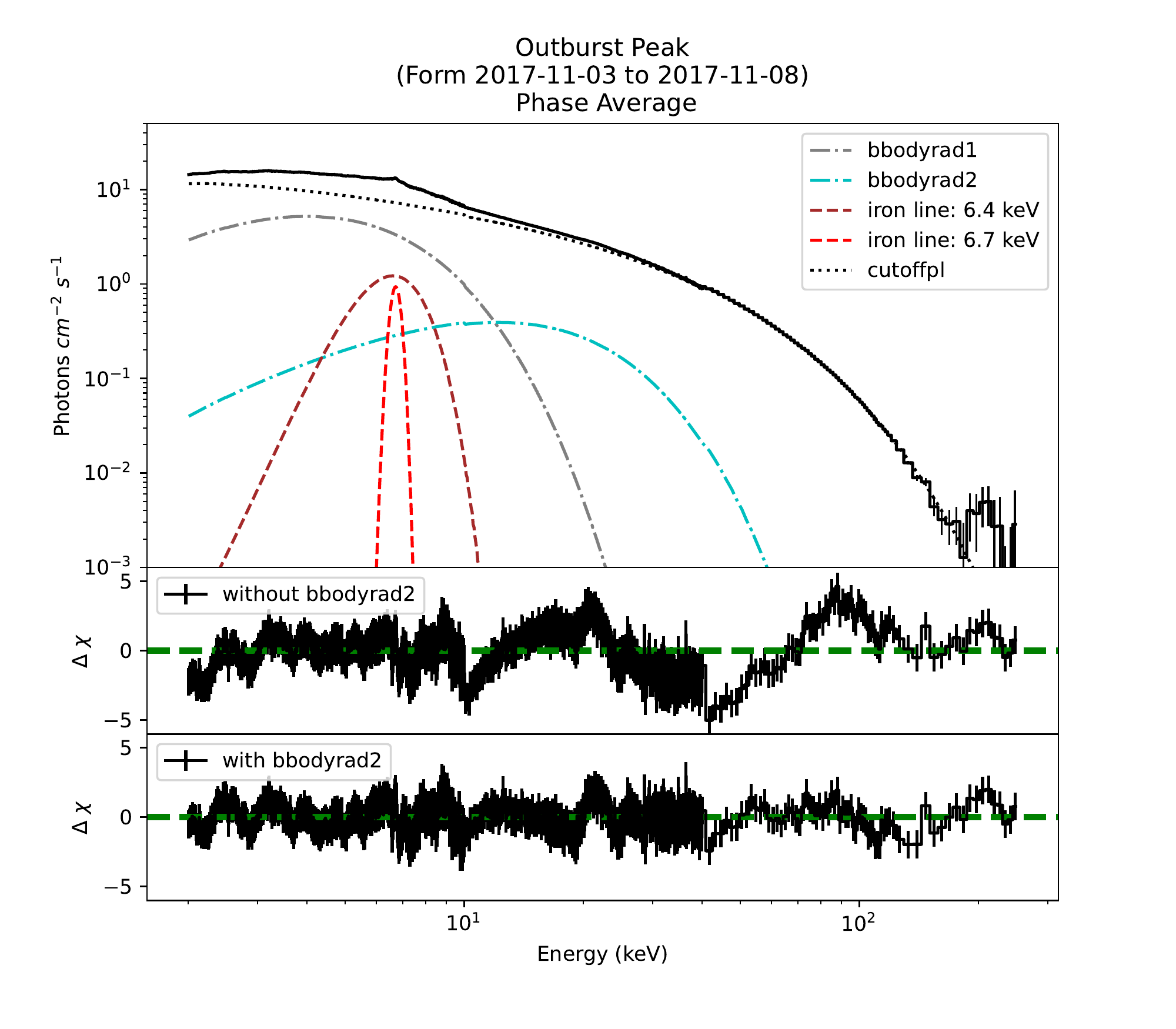}
    \centering\includegraphics[width=0.49\textwidth]{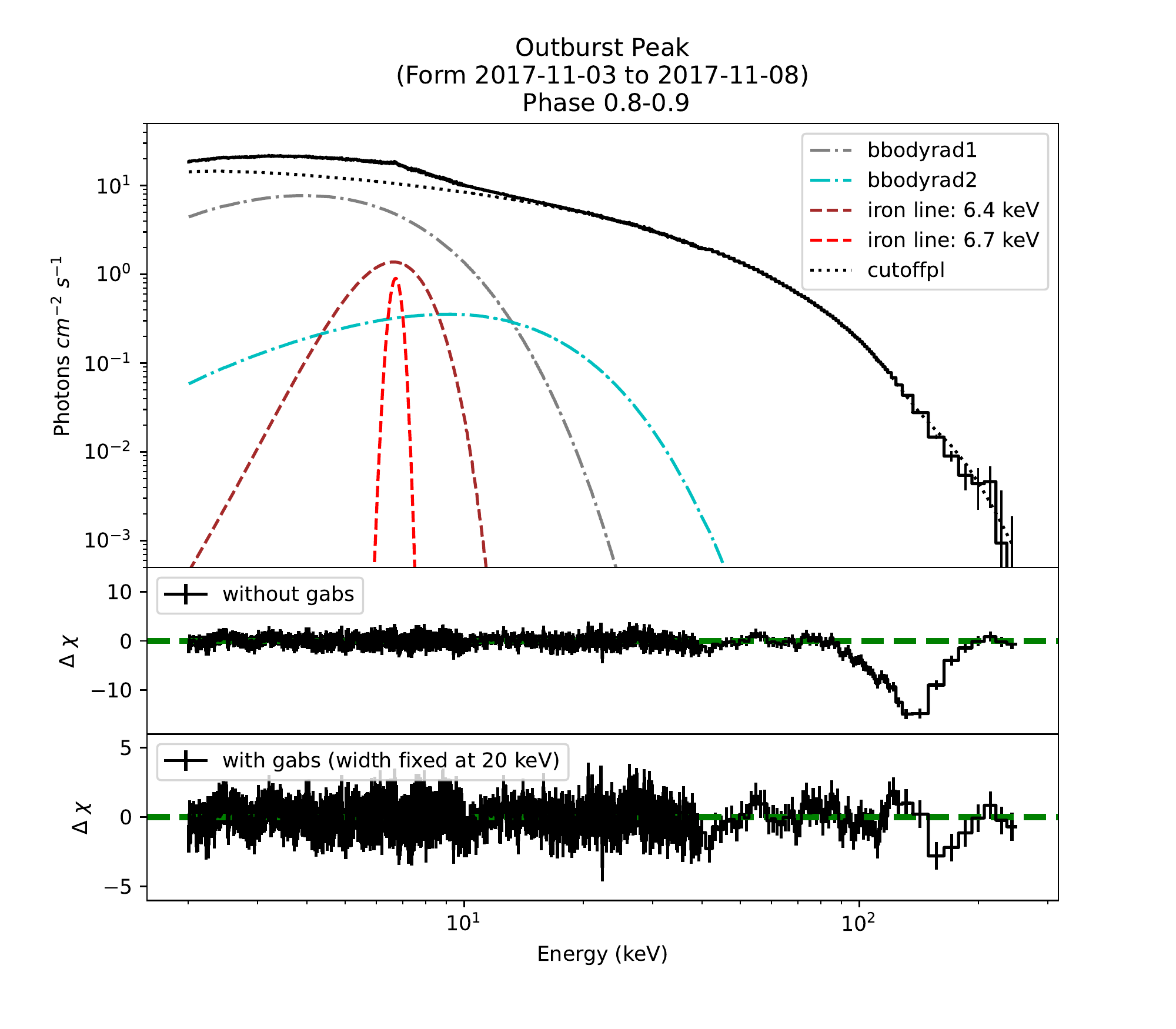}
    \caption{We combine the observations from Nov 3 to Nov 8 in 2017 for spectral fittings.
    Left panel: Spectral fitting and residuals of the phase-averaged spectrum.
    Right panel: Spectral fitting and residuals of the spectrum at phase $0.8-0.9$.}
    \label{spectra}
\end{figure}

\begin{figure}
    \centering\includegraphics[width=0.49\textwidth,height=1.2\textwidth]{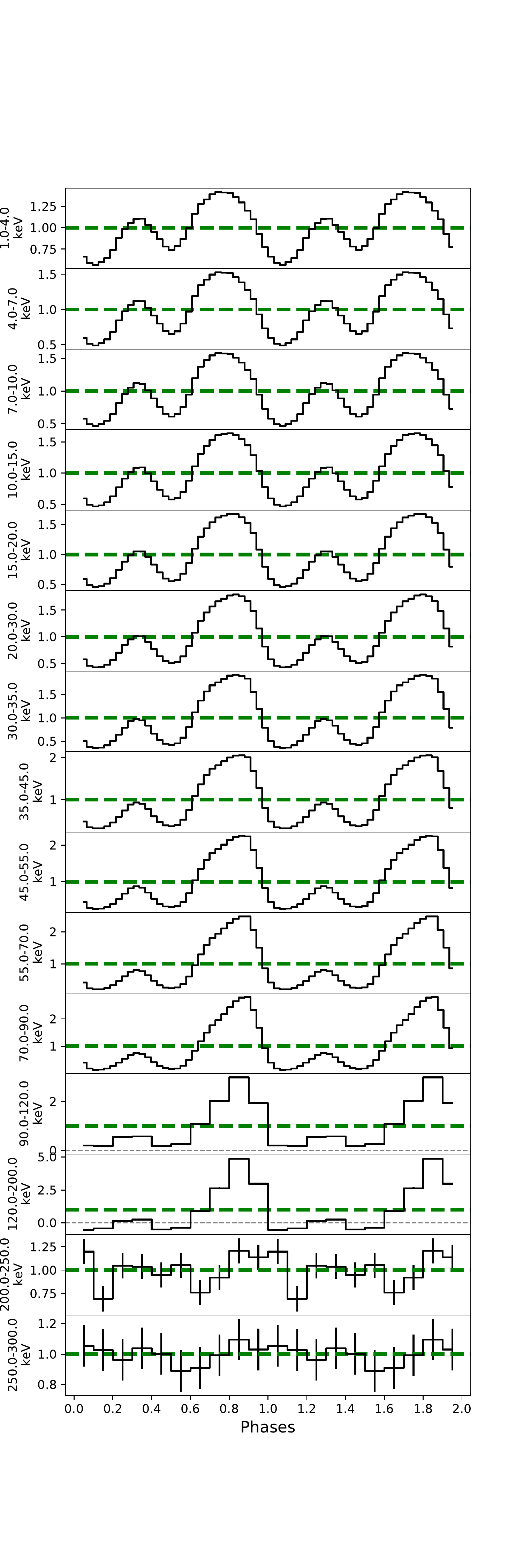}
    \centering\includegraphics[width=0.49\textwidth,height=1.2\textwidth]{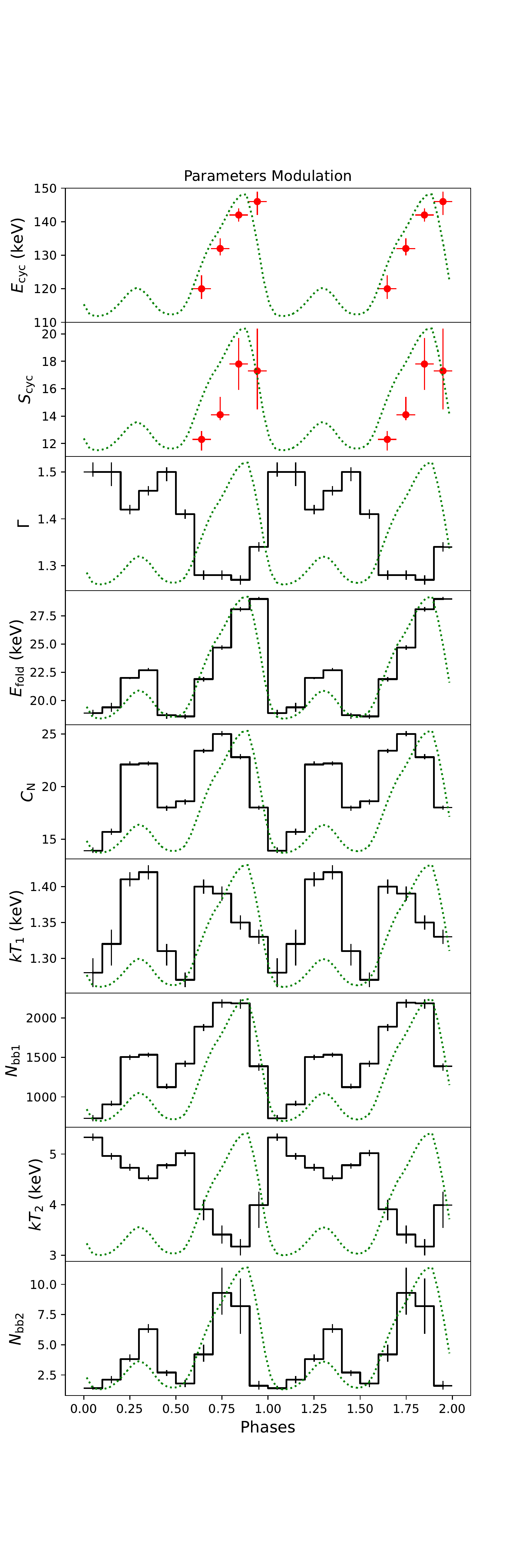}
    \caption{We combine the observations from Nov 3 to Nov 8 in 2017 for the pulse-profile analysis and the phase-resolved spectral analysis. 
    Left panel: Pulse profile evolves with energy. The background-subtracted pulse profiles are normalized by the mean count rate. The green dashed line is at 1.0, and the grey dotted line is at 0.
    Right panel: Fitting results of the phase-resolved spectral analysis. The red points show the parameters of the CRSF.}
    \label{phase-resolved}
\end{figure}

Our approach is to fit the phase-averaged spectrum and the phase-resolved spectra with the same spectral model, in order to clearly reveal the phase-dependence of the spectral features. We thus first fit the phase-averaged spectrum with the model \textit{TBabs}$\times$\textit{(bbodyrad1+cutoffpl+gaussian1+gaussian2)}. A single black body with a non-thermal component model and multiple iron emission lines can well describe the phase-averaged spectra in \cite{Kong2020ApJ}. The Tuebingen-Boulder ISM absorption model \textit{TBabs} is taken into account (\citealp{Wilms2000}), and the equivalent hydrogen column density $n_{\rm H}$ is fixed at $0.7 \times 10^{22}$ atom cm$^{-2}$, which is estimated through the Galactic H~\uppercase\expandafter{\romannumeral1} density in the direction of Swift J0243.6+6124 (\citealp{Kalberla2005}). 
The \textit{cutoffpl} is a simple non-thermal continuum with just three free parameters,
normalization coefficient ($C_{\rm N}$), photon index ($\Gamma$), and exponential folding energy ($E_{\rm fold}$), respectively.
We use two \textit{gaussian} models to fit multiple iron emission structures which were reported in \cite{Jaisawal2019}, and we fix the energies at 6.4 keV and 6.7 keV for neutral and ionized iron lines.
The black body \textit{bbodyrad1} with $kT_1\sim1.5$ keV and $N_{\rm bb1}\sim 2000$ were reported in \cite{Kong2020ApJ}.
The fitting reuslts for the phase-averaged spectrum are shown in the left panel of Figure~\ref{spectra}.
From the residuals, adding another black body \textit{bbodyrad2} with high $kT_2\sim4$ keV and small $N_{\rm bb2}\sim5$ can improve the $\chi^2$ via suppressing residuals. 
This structure has also proved necessary in \cite{Kong2020ApJ} and \cite{Tao2019}. Therefore, an extra component bbodyrad2 is added to the model to fit the phase-averaged spectrum.
No significant absorption feature in the phase-averaged spectrum can stimulate us to add a CRSF.

Next, the same model 
is used for the spectral fittings of ten different phase intervals. 
The fitting parameters are listed in Table~\ref{spectral_fitting}. 
The spectra in phase intervals in $0.0-0.6$ can be well fitted with this model, and the value of $\chi^2$ approaches to the degree of freedom (DOF). 
In contrast, the fitting residuals of the phase intervals in $0.6-1.0$ show obvious absorption features above 100 keV, which can be suppressed by introducing a gaussian absorption model \textit{gabs}; the parameters are listed in Table~\ref{spectral_fitting}. 
First the width $\sigma$ of the \textit{gabs} model is fixed at 20 keV, considering the low counting statics in high-energy bands ($>150$ keV); otherwise either the line width $\sigma$ or the line energy $E_{\rm cyc}$ can not be well constrained because these two parameters are coupled with each other. We note here that if the line width is set to another value such as 30 keV, the significance of the absorption component and the fit to the continuum spectrum do not change obviously. 
The fitting parameters in different phase intervals are plotted in the right panel of Figure~\ref{phase-resolved}.
The red points denote the line energy $E_{\rm cyc}$ and absorption strength $S_{\rm cyc}$ in the \textit{gabs} model.
We also find that $E_{\rm cyc}$ is phase dependent and varies between $120_{-3}^{+4}$ keV to $146_{-4}^{+3}$ keV along the pulse phase in $0.6-1.0$. 
The absorption feature has the maximum $E_{\rm cyc}=146_{-4}^{+3}$ keV and largest $S_{\rm cyc}=17.3_{-2.8}^{+3.1}$ ($6.3\ \sigma$ detection significance) at the pulse peak in phase interval. The detection significance of this CRSF varies between 6 to 18 $\sigma$ at other phases.
We plot the fitting and residuals in phase interval $0.8-0.9$, where the residuals show a significant absorption structure above 100 keV after removing the \textit{gabs} model from the overall model.

Now we free the width of the CRSF, which is fixed in the above spectral analyses; the results are listed in Table~\ref{free_width}. The CRSF’s width can be limited to $20-30$ keV, and such a wide width of CRSF strongly disfavors the possibility of a proton cyclotron feature.
We simulate 100,000 spectra using the best-fitting model parameters for phase 0.8-0.9 in Table~\ref{free_width} but with the CRSF excluded, and then fit these simulated spectra using the same model with and without an CRSF. The maximum $\Delta\chi^2$ is only 27, to be compared with $\Delta\chi^2$ = 336 obtained by fitting to the observed spectrum; this confirms that the CRSF is detected with high statistical significance.

\begin{figure}
    \centering\includegraphics{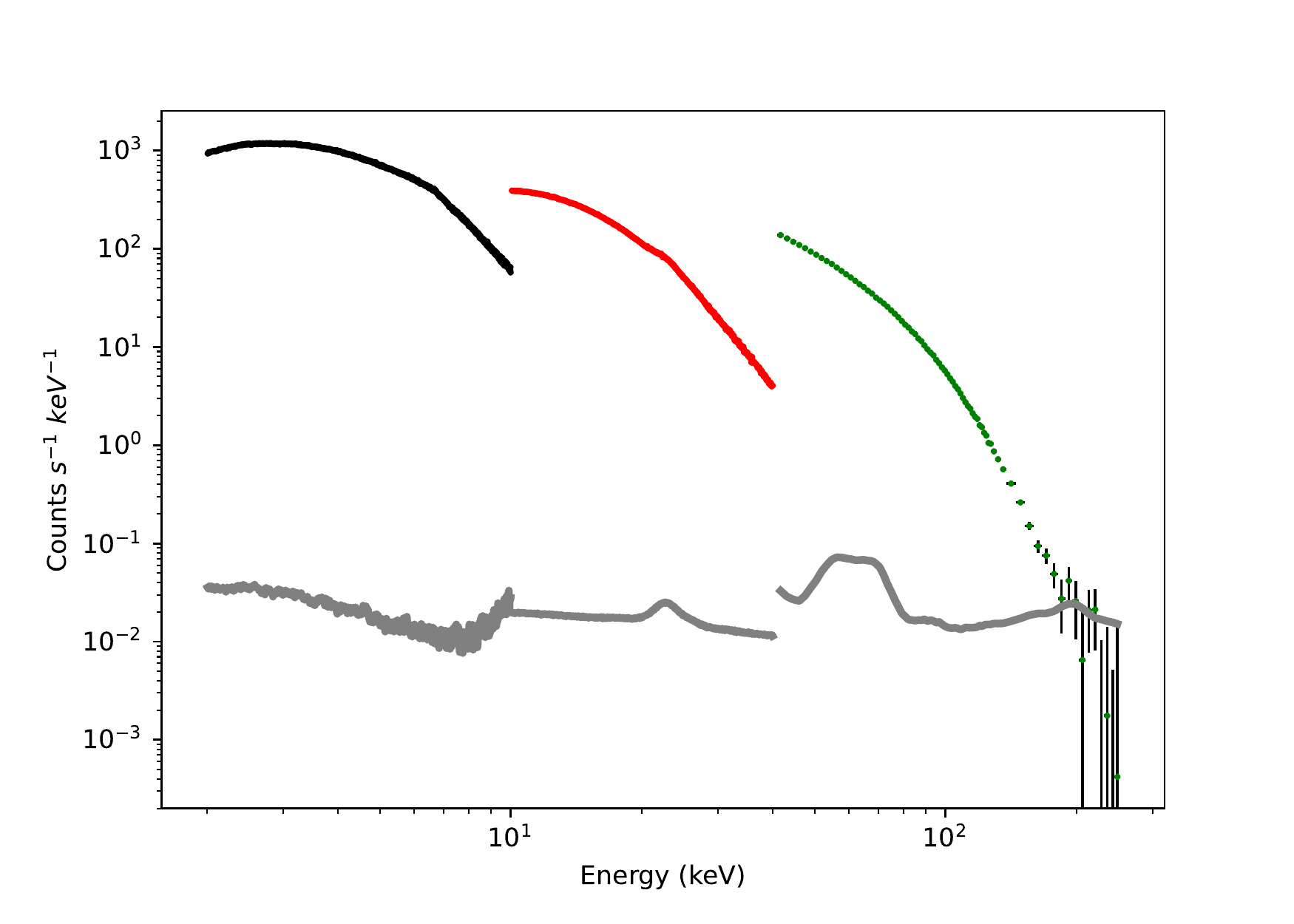}
    \caption{The net flux of LE, ME, and HE are shown in black, red, and green points, respectively. The flux and background are obtained from the phase-resolved spectrum in phase interval $0.8-0.9$. The gray line shows the systematic errors of the background flux; for the exposure time of $\sim 20$ ks, we choose the systematic error of 2.5\%, 1.5\%, and 1.5\% for LE, ME, and HE, respectively, based on the HXMT background models (\citealp{Liao2020a,Guo2020JHEAp,Liao2020b}). For such a long exposure, the statistical errors of the background flux are completely negligible.}
    \label{back_plot}
\end{figure}

From the phase-averaged and phase-resolved spectral fitting, the residuals do not support the addition of a $60-70$ keV line during the spectral fittings. 
We try to add a gabs line (fixed at Ecyc=70 keV, width=15 keV) during the fitting, and get a strength $<7\times10^{-16}$, thus disfavoring the existence of such an absorption feature. This proves that the $120-146$ keV absorption feature is the fundamental CRSF of Swift~J0243.6+6124, which is also the highest energy CRSF discovered so far.
In Figure~\ref{back_plot}, we show the background-subtracted flux (source flux) and the systematic error of the background flux for LE, ME, and HE; their statistical errors are completely negligible. It is clear that the background uncertainties begin to dominate only above 200 keV; therefore, our detection of the CRSF is not influenced by background uncertainties.

From Figure~\ref{phase-resolved}, the non-thermal component and two black body components also show significant phase-dependent structures. 
$\Gamma$ shows a clear bi-modal distribution: smaller than 1.3 during the main pulse and greater than 1.4 elsewhere, implying a harder spectrum during the main pulse. $E_{\rm fold}$ increases from $\sim18.6$ keV to $\sim 30$ keV when the phase changes from 0.5 to 1.0.
The two thermal components are also phase-dependent. 
The temperatures of \textit{bbodyrad1} and \textit{bbodyrad2} show a different correlation with the intensity in each phase. 
The negative correlation between $kT_2$ and intensity leads to a lower temperature $\sim 3$ keV for the main pulse in phase region $0.6-1.0$.

In addition, we also tested if other continuum models, such as \textit{highecut$\times$powerlaw}, \textit{NPEX} and \textit{fdcut}, may influence the absorption structure.
We found that the absorption structure around 140 keV is unchanged using different continuum models.
We also note that when the model is NPEX, the high-temperature blackbody component (bbodyrad2) is no longer required in the fitting.

\begin{sidewaystable}[ptbptbptb]
\begin{center}
\caption{Parameters of the spectral fitting at different phases.}
\resizebox{.95\columnwidth}{!}{
\begin{tabular}{cccccccccccccccc}
\hline
\hline
phase & & $0.0-0.1$ & $0.1-0.2$ & $0.2-0.3$ & $0.3-0.4$ & $0.4-0.5$ & $0.5-0.6$ & $0.6-0.7$ & $0.6-0.7$ & $0.7-0.8$ & $0.7-0.8$ & $0.8-0.9$ & $0.8-0.9$ & $0.9-1.0$ & $0.9-1.0$
\\
\hline
TBabs & $n_{\rm H}\ (10^{22}\ \rm cm^{-2})$ & $0.7$ (fixed) & $0.7$ (fixed) & $0.7$ (fixed) & $0.7$ (fixed) & $0.7$ (fixed) & $0.7$ (fixed) & $0.7$ (fixed) & $0.7$ (fixed) & $0.7$ (fixed) & $0.7$ (fixed) & $0.7$ (fixed) & $0.7$ (fixed) & $0.7$ (fixed) & $0.7$ (fixed)
\\
gabs & $E_{\rm cyc}$ (keV) & ... & ... & ... & ... & ... & ... & ... & $120_{-3}^{+4}$ & ... & $132_{-2}^{+3}$ & ... & $142_{-2}^{+2}$ & ... &$146_{-4}^{+3}$
\\
& $\sigma_{\rm cyc}$ (keV) & ... & ... & ... & ... & ... & ... & ... & $20$ (fixed) & ... & $20$ (fixed) & ... & $20$ (fixed) & ... & $20$ (fixed)
\\
& $S_{\rm cyc}$ & ... & ... & ... & ... & ... & ... & ... & $12.3_{-0.8}^{+0.6}$ & ... & $14.1_{-0.4}^{+1.3}$ & ... & $17.8_{-1.9}^{+1.9}$ & ... & $17.3_{-2.8}^{+3.1}$
\\
gaussian1 & $E_{\rm Fe1}$ (keV) & 6.4 (fixed) & 6.4 (fixed) & 6.4 (fixed) & 6.4 (fixed) & 6.4 (fixed) & 6.4 (fixed) & 6.4 (fixed) & 6.4 (fixed) & 6.4 (fixed) & 6.4 (fixed) & 6.4 (fixed) & 6.4 (fixed) & 6.4 (fixed) & 6.4 (fixed)
\\
& $\sigma_{\rm Fe1}$ (keV) & $1.00_{-0.03}^{+0.05}$ & $0.95_{-0.02}^{+0.05}$ & $0.84_{-0.07}^{+0.05}$ & $0.88_{-0.05}^{+0.05}$ & $0.90_{-0.03}^{+0.04}$ & $1.05_{-0.04}^{+0.06}$ & $1.14_{-0.04}^{+0.05}$ & $1.00_{-0.07}^{+0.08}$ & $1.19_{-0.06}^{+0.04}$ & $1.00_{-0.07}^{+0.05}$ & $1.20_{-0.05}^{+0.05}$ & $1.10_{-0.08}^{+0.08}$ & $1.06_{-0.08}^{+0.05}$ & $1.01_{-0.08}^{+0.07}$
\\
& norm1 & $0.39_{-0.01}^{+0.02}$ & $0.37_{-0.03}^{+0.03}$ & $0.35_{-0.03}^{+0.03}$ & $0.34_{-0.03}^{+0.02}$ & $0.43_{-0.01}^{+0.03}$ & $0.55_{-0.03}^{+0.03}$ & $0.73_{-0.06}^{+0.07}$ & $0.49_{-0.05}^{+0.05}$ & $0.85_{-0.04}^{+0.05}$ & $0.46_{-0.04}^{+0.03}$ & $0.88_{-0.07}^{+0.06}$ & $0.55_{-0.06}^{+0.06}$ & $0.52_{-0.06}^{+0.04}$ & $0.44_{-0.04}^{+0.04}$
\\
gaussian2 & $E_{\rm Fe2}$ (keV) & 6.7 (fixed) & 6.7 (fixed) & 6.7 (fixed) & 6.7 (fixed) & 6.7 (fixed) & 6.7 (fixed) & 6.7 (fixed) & 6.7 (fixed) & 6.7 (fixed) & 6.7 (fixed) & 6.7 (fixed) & 6.7 (fixed) & 6.7 (fixed) & 6.7 (fixed)
\\
& $\sigma_{\rm Fe2}$ (keV) & $0.17_{-0.02}^{+0.03}$ & $0.17_{-0.02}^{+0.03}$ & $0.18_{-0.03}^{+0.03}$ & $0.19_{-0.04}^{+0.04}$ & $0.15_{-0.03}^{+0.03}$ & $0.15_{-0.03}^{+0.03}$ & $0.17_{-0.03}^{+0.03}$ & $0.18_{-0.05}^{+0.05}$ & $0.16_{-0.03}^{+0.03}$ & $0.18_{-0.04}^{+0.03}$ & $0.17_{-0.02}^{+0.04}$ & $0.20_{-0.03}^{+0.04}$ & $0.17_{-0.03}^{+0.03}$ & $0.18_{-0.03}^{+0.02}$
\\
& norm2 ($10^{-2}$)& $5.1_{-0.5}^{+0.8}$ & $5.4_{-0.4}^{+0.6}$ & $5.2_{-0.9}^{+0.7}$ & $6.1_{-1.2}^{+1.2}$ & $5.5_{-0.8}^{+0.7}$ & $5.6_{-0.7}^{+1.1}$ & $5.6_{-0.9}^{+1.1}$ & $5.8_{-1.2}^{+1.1}$ & $5.5_{-1.3}^{+1.4}$ & $6.3_{-1.0}^{+0.6}$ & $5.4_{-0.7}^{+1.1}$ & $6.4_{-1.1}^{+1.3}$ & $5.5_{-0.7}^{+1.0}$ & $5.6_{-1.1}^{+0.8}$
\\
bbodyrad1 & $kT_1$ (keV) & $1.28_{-0.02}^{+0.02}$ & $1.32_{-0.03}^{+0.02}$ & $1.41_{-0.01}^{+0.01}$ & $1.42_{-0.01}^{+0.01}$ & $1.31_{-0.02}^{+0.01}$ & $1.27_{-0.01}^{+0.01}$ & $1.22_{-0.03}^{+0.02}$ & $1.40_{-0.01}^{+0.01}$ & $1.16_{-0.02}^{+0.01}$ & $1.39_{-0.01}^{+0.01}$ & $1.13_{-0.01}^{+0.01}$ & $1.35_{-0.01}^{+0.01}$  & $1.25_{-0.02}^{+0.02}$ & $1.33_{-0.01}^{+0.01}$
\\
& $N_{\rm bb1}$ & $730_{-36}^{+41}$ & $907_{-22}^{+46}$ &$1505_{-38}^{+28}$ & $1534_{-29}^{+27}$ & $1124_{-21}^{+42}$ & $1420_{-42}^{+40}$ & $2658_{-134}^{+226}$ & $1887_{-53}^{+35}$ & $3649_{-144}^{+198}$ & $2194_{-63}^{+43}$ & $3695_{-219}^{+157}$ & $2185_{-68}^{+49}$ & $1680_{-105}^{+87}$ & $1388_{-56}^{+34}$
\\
bbodyrad2 & $kT_2$ (keV) & $5.33_{-0.07}^{+0.08}$ & $4.96_{-0.07}^{+0.06}$ & $4.73_{-0.06}^{+0.08}$ & $4.52_{-0.05}^{+0.06}$ & $4.78_{-0.07}^{+0.04}$ & $5.02_{-0.06}^{+0.06}$ & $2.12_{-0.03}^{+0.07}$ & $3.91_{-0.22}^{+0.19}$ & $2.19_{-0.05}^{+0.03}$ & $3.41_{-0.18}^{+0.18}$ & $2.12_{-0.05}^{+0.04}$ & $3.17_{-0.17}^{+0.15}$ & $2.30_{-0.11}^{+0.06}$ & $3.99_{-0.45}^{+0.27}$
\\
& $N_{\rm bb2}$ & $1.4_{-0.1}^{+0.2}$ & $2.1_{-0.3}^{+0.3}$ & $3.8_{-0.2}^{+0.4}$ & $6.3_{-0.3}^{+0.4}$ & $2.7_{-0.3}^{+0.2}$ & $1.8_{-0.3}^{+0.3}$ & $107.7_{-19.5}^{+15.7}$ & $4.2_{-0.6}^{+0.8}$ & $147.6_{-14.1}^{+26.4}$ & $9.3_{-1.8}^{+2.1}$ & $147.2_{-15.1}^{+15.4}$ & $8.2_{-2.3}^{+2.3}$ & $24.3_{-5.7}^{+6.3}$ & $1.6_{-0.3}^{+0.4}$
\\
cutoffPL & $\Gamma$ & $1.50_{-0.01}^{+0.02}$ & $1.50_{-0.03}^{+0.02}$ & $1.42_{-0.01}^{+0.01}$ & $1.46_{-0.01}^{+0.01}$ & $1.50_{-0.02}^{+0.01}$ & $1.41_{-0.01}^{+0.01}$ & $1.21_{-0.01}^{+0.01}$ & $1.28_{-0.01}^{+0.01}$ & $1.20_{-0.01}^{+0.01}$ & $1.28_{-0.01}^{+0.01}$ & $1.19_{-0.01}^{+0.01}$ & $1.27_{-0.01}^{+0.01}$ & $1.31_{-0.01}^{+0.01}$ & $1.34_{-0.01}^{+0.01}$
\\
& $E_{\rm fold}$ (keV) & $18.9_{-0.2}^{+0.3}$ & $19.4_{-0.4}^{+0.4}$ & $22.0_{-0.1}^{+0.2}$ & $22.7_{-0.1}^{+0.2}$ & $18.7_{-0.3}^{+0.2}$ & $18.6_{-0.2}^{+0.2}$ & $20.9_{-0.1}^{+0.1}$ & $21.9_{-0.2}^{+0.2}$ & $23.4_{-0.1}^{+0.1}$ & $24.7_{-0.2}^{+0.2}$ & $26.5_{-0.1}^{+0.1}$ & $28.1_{-0.2}^{+0.2}$ & $28.1_{-0.1}^{+0.1}$ & $29.0_{-0.1}^{+0.2}$
\\
& $C_{\rm N}$ & $13.9_{-0.2}^{+0.3}$ & $15.7_{-0.3}^{+0.3}$ & $22.1_{-0.1}^{+0.3}$ & $22.2_{-0.2}^{+0.2}$ & $18.0_{-0.3}^{+0.2}$ & $18.6_{-0.3}^{+0.2}$ & $21.1_{-0.5}^{+0.2}$ & $23.4_{-0.2}^{+0.2}$ & $21.2_{-0.4}^{+0.4}$ & $25.0_{-0.2}^{+0.3}$ & $19.0_{-0.3}^{+0.5}$ & $22.8_{-0.2}^{+0.3}$ & $17.0_{-0.2}^{+0.3}$ & $18.0_{-0.2}^{+0.2}$
\\
Fitting & Reduced-$\chi^{2}$ (dof) & 1.08 (1585) & 0.98 (1588) & 0.93 (1588) & 0.94 (1588) & 1.01 (1588) & 1.00 (1588) & 1.13 (1588) & 0.85 (1586) & 1.14 (1588) & 0.83 (1586) & 1.23 (1588) & 0.81 (1586) & 1.06 (1588) & 0.87 (1586)
\\
\hline
\hline
\end{tabular}
\label{spectral_fitting}
}    
\begin{list}{}{}
    \item[Note]{: Uncertainties are reported at the 90\% confidence interval and computed using MCMC (Markov Chain Monte Carlo) of length 10,000. The 1\%, 0.5\%, 0.5\% system error for LE, ME, and HE have been added during spectral fittings.}
\end{list}
\end{center}
\end{sidewaystable}

\begin{sidewaystable}[ptbptbptb]
\begin{center}
\caption{The width of CRSF lines set free to change}
\resizebox{.95\columnwidth}{!}{
\begin{tabular}{ccccccc}
\hline
\hline
phase & & $0.6-0.7$ & $0.7-0.8$ & $0.8-0.9$ & $0.9-1.0$ 
\\
\hline
TBabs & $n_{\rm H}\ (10^{22}\ \rm cm^{-2})$ & $0.7$ (fixed) & $0.7$ (fixed) & $0.7$ (fixed) & $0.7$ (fixed) 
\\
gabs & $E_{\rm cyc}$ (keV) & $120_{-2}^{+8}$ & $133_{-2}^{+7}$ & $157_{-17}^{+2}$ & $165_{-20}^{+1}$
\\
& $\sigma_{\rm cyc}$ (keV) & $21_{-1}^{+6}$ & $21_{-1}^{+5}$ & $29_{-5}^{+1}$ & $29_{-7}^{+1}$
\\
& $S_{\rm cyc}$ & $13_{-1}^{+10}$ & $15_{-1}^{+7}$ & $35_{-8}^{+4}$ & $37_{-16}^{+3}$
\\
gaussian1 & $E_{\rm Fe1}$ (keV) & 6.4 (fixed) & 6.4 (fixed) & 6.4 (fixed) & 6.4 (fixed) 
\\
& $\sigma_{\rm Fe1}$ (keV) & $1.00_{-0.09}^{+0.05}$ & $1.01_{-0.02}^{+0.06}$ & $1.08_{-0.05}^{+0.05}$ & $1.01_{-0.05}^{+0.05}$
\\
& norm1 & $0.49_{-0.06}^{+0.02}$ & $0.46_{-0.04}^{+0.05}$ & $0.53_{-0.07}^{+0.02}$ & $0.44_{-0.03}^{+0.04}$
\\
gaussian2 & $E_{\rm Fe2}$ (keV) & 6.7 (fixed) & 6.7 (fixed) & 6.7 (fixed) & 6.7 (fixed)
\\
& $\sigma_{\rm Fe2}$ (keV) & $0.18_{-0.05}^{+0.04}$ & $0.18_{-0.03}^{+0.06}$ & $0.20_{-0.03}^{+0.06}$ & $0.18_{-0.03}^{+0.02}$
\\
& norm2 ($10^{-2}$) & $5.8_{-1.2}^{+1.2}$ & $6.3_{-1.1}^{+1.8}$ & $6.5_{-0.8}^{+2.1}$ & $5.6_{-1.1}^{+0.9}$
\\
bbodyrad1 & $kT_1$ (keV) & $1.40_{-0.01}^{+0.02}$ & $1.39_{-0.01}^{+0.01}$ & $1.36_{-0.01}^{+0.01}$ & $1.34_{-0.01}^{+0.01}$
\\
& $N_{\rm bb1}$ & $1887_{-37}^{+40}$ & $2195_{-40}^{+85}$ & $2151_{-25}^{+58}$ & $1394_{-15}^{+45}$
\\
bbodyrad2 & $kT_2$ (keV) & $3.9_{-0.2}^{+0.3}$ & $3.4_{-0.1}^{+0.2}$ & $3.4_{-0.2}^{+0.2}$ & $4.0_{-0.3}^{+0.5}$
\\
& $N_{\rm bb2}$ & $4.2_{-0.4}^{+0.9}$ & $9.4_{-1.2}^{+2.1}$ & $6.4_{-1.0}^{+1.8}$ & $1.8_{-0.4}^{+0.3}$
\\
cutoffPL & $\Gamma$ & $1.28_{-0.01}^{+0.02}$ & $1.28_{-0.01}^{+0.01}$ & $1.28_{-0.01}^{+0.01}$ & $1.34_{-0.01}^{+0.01}$
\\
& $E_{\rm fold}$ (keV) & $21.9_{-0.1}^{+0.7}$ & $24.7_{-0.1}^{+0.4}$ & $28.5_{-0.2}^{+0.3}$ & $29.2_{-0.4}^{+0.1}$
\\
& $C_{\rm N}$ & $23.4_{-0.2}^{+0.4}$ & $25.0_{-0.4}^{+0.1}$ & $22.9_{-0.2}^{+0.2}$ & $18.0_{-0.2}^{+0.1}$
\\
Fitting & Reduced-$\chi^{2}$ (dof) & 0.85 (1585) & 0.82 (1585) & 0.80 (1585) & 0.86 (1585) 
\\
\hline
\hline
\end{tabular}
\label{free_width}
}    
\begin{list}{}{}
    \item[Note]{: Uncertainties are reported at the 90\% confidence interval and computed using MCMC (Markov Chain Monte Carlo) of length 10,000. The 1\%, 0.5\%, 0.5\% system error for LE, ME, and HE have been added during spectral fittings.}
\end{list}
\end{center}
\end{sidewaystable}

\section{Discussion}

We have analyzed the data from the high cadence and broad band \emph{Insight}-HXMT observations of the 2017-2018 outburst of the first Galactic ULX Swift~J0243.6+6124. 
A CRSF is discovered with high statistical significance at energies around $120-146$ keV, in a period when the source stayed at a luminosity around $2\times10^{39}$ erg s$^{-1}$, far beyond the super-critical luminosity of the accretion column and the Eddington luminosity of the accreting NS. 
The energy of this newly detected CRSF is much higher than both the $\sim90$ keV fundamental CRSF of GRO~J1008-57 (\citealp{Ge2020ApJ}) and the $\sim100$ keV first harmonic CRSF of 1A~0526+261 (\citealp{Kong2021ApJL, Kong2022arXiv220411222K});  both were the previous high energy records of CRSF detection, whose existence was firmly established for the first time by \emph{Insight}-HXMT with a significance much higher than $5\ \sigma$. 
We note that a marginal detection (absorption depth unconstrained) of the possible third CRSF harmonic at $\sim$128 keV was reported in the spectrum of MAXI J1409-619 (\citealp{Orlandini2012ApJ}).
The discovery of the CRSF from Swift~J 0243.6+6124 reported here thus serves as the highest energy CRSF detected so far from a NS harbored in an accreting X-ray binary system. 
The strength of the magnetic field can be calculated with the following relation: 
\begin{equation}
    E_{\rm cyc}\ =\ \frac{n}{(1+z)}\frac{\hbar e B}{m_{\rm e} c}\ \approx\  11.6\ \frac{n}{(1+z)} \times B_{12}\ (\rm keV),
\end{equation}
where $B_{12}$ is the magnetic field strength in units of $10^{12}$ G, and $z=(1-2GM/r_{\rm cyc}c^2)^{-1/2}-1$ is the gravitational redshift due to the NS mass and height of the line forming region ($r_{\rm cyc}$). $n$ is the number of the Landau level involved: e.g., $n\ =\ 1$ for the fundamental line and $n\ \ge \ 2$ for its harmonics.
The above equation assumes that the absorption takes place at the NS surface. 
However, in reality the absorption can only happen somewhere above the NS surface, and the higher the energy, the closer the absorption takes place to the NS surface.  
We thus take $E_{\rm cyc}\sim 146$ keV and $z=0.25$ for the emission from the surface of a typical NS ($M=1.4 M_{\odot}$ and $R=10$ km) in the equation and obtain a lower limit of magnetic field $B\sim1.6\times10^{13}$ G.
This is the strongest surface magnetic field of a NS measured directly with an electron CRSF.
Here, we also consider the attenuation of the magnetic field with height, for the dipole magnetic field $B\varpropto r^{-3}$, and for the quadrupole magnetic field $B\varpropto r^{-4}$. We find that compared with the influence of redshift on cyclotron energy, the influence of height change on CRSF energy is more significant, which further shows that $B\sim1.6\times10^{13}$ G represents a lower limit of the magnetic field on the NS's surface.

Our result appears to be at odds with the comparatively small magnetospheric radius estimated for the source by several authors independently based on completely different observables \citep{Jaisawal2019, Tao2019, Eijnden2019MNRAS.487.4355V, Doroshenko2020MNRAS}. 
In particular, the non-transition of the source to a ``propeller'' state in quiescence implies an upper limit on the dipole field of $\le3\times10^{12}$\,G, i.e. an order of magnitude lower that given by the CRSF measurement. It is, therefore, necessary to discuss the origin of this discrepancy. 
First of all, it is important to emphasize that the evidence for the transition of the accretion disk to the radiative-pressure dominated state close to the peak of the outburst appears to be unambiguously supported by several arguments, and requires that the magnetosphere remains comparatively compact \citep{Doroshenko2020MNRAS}. 
In fact, as discussed in the latter paper, standard scaling of the magnetosphere size with flux implies a good agreement between estimates obtained at the peak and at the very end of the outburst, i.e. within the range of over five orders of magnitude in luminosity. 
It is, therefore, reasonable to assume that the magnetosphere is indeed rather compact. 
On the other hand, as discussed by \cite{Doroshenko2020MNRAS}, the field implied by such a compact magnetosphere ($\sim2\times10^{7}$\,cm close to the peak of the outburst) is lower than required to explain a dramatic change of the pulse profile around either $L_x\sim1.5\times10^{38}$\,erg\,s$^{-1}$ interpreted by \cite{Doroshenko2020MNRAS}, or around $L_x\sim4.4\times10^{38}$\,erg\,s$^{-1}$ interpreted by \cite{Kong2020ApJ} as the onset of the accretion column.

Indeed, the critical luminosity for a NS XRB system corresponding to such transition and switch of the emission pattern from ``pencil" to ``fan" beams is connected to the magnetic field in the following way with typical NS parameters of \emph{M} = 1.4 $M_{\odot}$, \emph{R} = 10 km, $\Lambda$ = 0.1, and $\omega$ = 1 (\citealp{Becker2012}): 
\begin{equation}
L_{\rm crit} = 1.5 \times L_{37} B_{12}^{16/15},
\end{equation}
where $L_{37}$ is the luminosity in units of $10^{37}$ erg s$^{-1}$. 
Here for Swift~J0243.6+6124, a critical luminosity of $L_{\rm crit}=2.9\times10^{38}$ erg s$^{-1}$ would be expected under a magnetic field of $1.6\times10^{13}$ G, i.e. in perfect agreement with our estimate based on the observed CRSF energy, and significantly higher than could be expected for a $\sim10^{12}$\,G field implied by other estimates. 
This fact was actually noted by \cite{Doroshenko2020MNRAS} who argued that this discrepancy can be explained either by a small value of the scaling factor $k=R_{m}/R_{A}$ (where $R_{A}$ is standard equation for Alfven radius), or the presence of the strong multipole components. 
Our detection of the CRSF thus strongly implies that the latter scenario is indeed correct, which is the first unambiguous evidence for the presence of the multipole fields in accreting neutron stars.

We note that a similar scenario has been invoked for several other pulsing ultraluminous X-ray sources (pULXs) including SMC X-3 (\citealp{Tsygankov2017AA}), GRO~J1744-28 (\citealp{2020A&A...643A..62D}), ULX M82 X-2 (\citealp{2021MNRAS.504..701B}), M51 ULX-8 (\citealp{2019MNRAS.486....2M}), and others (\citealp{2021MNRAS.504..701B}). Our result unambiguously confirms that this scenario is indeed viable and certainly realized in at least one pulsating ULX (i.e., Swift J0243.6+6124).

\section{Summary and Conclusions}

We report the phase-resolved analysis of the data of the first Galactic NS ULX Swift~J0243.6+6124 observed by \emph{Insight}-HXMT during its 2017-2018 outburst. 
The 120 to 146 keV CRSF is robustly detected in a spin phase region of $0.6-1.0$ at a significance level of 
$6-18\ \sigma$, when the source was at a luminosity of $\sim 2\times 10^{39}$ erg s$^{-1}$, far above its critical luminosity for the transition from a ``pencil" to a ``fan" emission pattern. 
This serves as the first unambiguous measurement of a CRSF from ULXs, and also the highest CRSF energy discovered so far from all neutron star X-ray binary systems. The measured surface magnetic field of $\sim1.6\times10^{13}$ G for Swift~J0243.6+6124 is the strongest for all known neutron stars with detected electron CRSFs, and also the strongest for all NS ULXs, unless the cases of marginal detection of possible proton CRSFs are confirmed with future instruments of a much larger effective area than the current instruments at 1-30 keV, e.g., the enhanced X-ray Timing and Polarimetry (eXTP) observatory to be launched in around 2027 (\citealp{Zhang2019SCPMAeXTP}).

Considering that several independent estimates previously constrained the dipole component of the magnetic field in Swift J0243.6+6124 to be an order of magnitude weaker, we conclude that the observed CRSF traces the multipole component of the field which, therefore, dominates the field in the vicinity of the neutron star's surface.
The presence of multipole field components has been previously argued for several other pULXs based on various arguments, however, our result represents the first and the unambiguous confirmation of this scenario.
We emphasize that the detection of a CRSF at such a high energy has only been made possible by the large effective area of HXMT in a hard X-ray band, and of course, a lucky coincidence that the mission was launched in time to observe the first and only giant outburst of Swift J0243.6+6124.
Our results are highly relevant for pULXs in general and allow us to settle several inconsistencies in the interpretation of pULX's phenomenology. 
In particular, the inconsistency between estimates arguing for a low field based on the spin-up rate and the fact that strong winds are apparently launched by most pULXs which would be impossible in the presence of magnetar fields (\citealp{King2019MNRAS}), and those arguing for ultra-strong fields based on the fact that only low plasma opacity realized in ultra-magnetized plasma might explain their observed luminosities (\citealp{2015MNRAS.454.2539M}).

\acknowledgments
We thanks the anonymous referee for valuable comments and suggestions. This work used data from the \emph{Insight}-HXMT mission, a project funded by China National Space Administration (CNSA) and the Chinese Academy of Sciences (CAS).
This work is supported by the National Key R\&D Program of China (2021YFA0718500) and the National Natural Science Foundation of China under grants U1838201, U2038101, U1838202, 11733009, 12122306, 12173103, U1838104, U1938101, U1938103, U2031205, and Guangdong Major Project of Basic and Applied Basic Research (Grant No. 2019B030302001). 

\bibliography{ref}
\bibliographystyle{aasjournal}

\end{document}